\documentclass[aps,pra,10pt, twocolumn, amsmath, amssymb, showpacs, superscriptaddress]{revtex4-1}
\usepackage{mathrsfs,amsfonts,graphicx,cancel,color,bm,ulem}
\usepackage[english]{babel}

\def\bra#1{\langle{#1}|}
\def\ket#1{|{#1}\rangle}
\def\braket#1{\langle{#1}\rangle}

{\catcode`\|=\active 
  \gdef\Braket#1{\begingroup
\mathcode`\|32768\let|\BraVert\left<{#1}\right>\endgroup}}
\def\BraVert{\egroup\,\mid\,\bgroup}

\definecolor{Blue}{rgb}{0,0,1}
\definecolor{Red}{rgb}{1,0,0}
\definecolor{Green}{rgb}{0,1,0}
\definecolor{Purp}{rgb}{.2,0,.2}
\definecolor{white}{rgb}{1,1,1}

\newcommand{\ep}{\epsilon}

\newcommand{\tr}{\mathop{\text{tr}}\nolimits}

\newtheorem{lemma}{Lemma}
\newtheorem{theorem}{Theorem}

\newcommand{\Proof}{\noindent\textbf{Proof.} } 
\newcommand{\Proofa}[1]{\noindent\textbf{Proof of #1.} } 
\newcommand{\qed}{$\square$}

\definecolor{dgreen}{rgb}{0,0.5,0}

\def\cO{${\cal O}$ }

\begin{document}

\date{\today}

\author{Rosario Fazio}
\affiliation{NEST, Scuola Normale Superiore and Istituto Nanoscienze-CNR, I-56126 Pisa, Italy}
\affiliation{Centre for Quantum Technologies, National University of Singapore, Singapore}

\author{Kavan Modi}
\affiliation{Department of Physics, University of Oxford, Clarendon Laboratory, Oxford OX1 3PU, UK}
\affiliation{Centre for Quantum Technologies, National University of Singapore, Singapore}

\author{Saverio Pascazio}
\affiliation{Dipartimento di Fisica and MECENAS, Universit\`a di Bari, I-70126 Bari, Italy}
\affiliation{INFN, Sezione di Bari, I-70126 Bari, Italy}

\author{Vlatko Vedral}
\affiliation{Department of Physics, University of Oxford, Clarendon Laboratory, Oxford OX1 3PU, UK}
\affiliation{Centre for Quantum Technologies, National University of Singapore, Singapore}
\affiliation{Department of Physics, National University of Singapore, Singapore}

\author{Kazuya Yuasa}
\affiliation{Department of Physics, Waseda University, Tokyo 169-8555, Japan}

\title{Witnessing the quantumness of a single system:\\ from anticommutators to interference and discord}

\begin{abstract}
We introduce a method to witness the quantumness of a system. The method relies on the fact that the anticommutator of two classical states is always positive. By contrast, we show that there is always a nonpositive anticommutator due to any two quantum states. We notice that interference depends on the trace of the anticommutator of two states and it is therefore operationally more suitable to detect quantumness by looking at anticommutators of states rather than their commutators. 
\end{abstract}


\maketitle

\section{Introduction}
\label{sec:intro}
What is the \textit{quantumness}  of a single physical system? This question, that goes back to the foundations of quantum mechanics, has become of ``practical" importance with the advent of quantum information processing. There are a number of tasks in computation and communication that can be performed only if quantum resources are available. This is the case, for example, when nonclassical states of light are employed in communication or metrology \cite{qlight}, or when mathematical models can be simplified beyond classical limits by tailored quantum systems \cite{occam}. It is therefore of paramount importance to put on firm quantitative terms the often elusive concept of quantumness. 

There are many different aspects of quantumness. For instance, quantumness is revealed in the form of nonlocality by the violation of Bell's inequality or in the form of contextuality by the Kochen-Specker test \cite{arXiv:1304.1292}. However, it involves two parties, and it is a quantumness that is manifested in the correlation between them. We are interested in the quantumness of a single system, that involves many aspects and approaches. Here, we shall focus on non-commutativity. It is usually stated that the main difference between quantum and classical physics is that quantum observables do not commute while all their classical counterparts do. In other words, some properties of quantum systems cannot be specified simultaneously. Well known examples are the position and the momentum of any quantum particle, or the $x$ and $z$ spin components of a spin-1/2 particle. However, quantum observables are only half of the picture. There are also quantum states (density operators). Classical states, on the other hand, are  simply-speaking probability distributions. In other words, all classical states commute with one another, while quantum states in general do not (quantum states, being Hermitian operators, are after all also observables). Can quantumness, therefore, be thought of as the degree of non-commutation of quantum states? How is it possible to quantify this concept in an operational way amenable of an experimental implementation?

In this Article we would like to relate the quantumness of a system to the non-commutativity of its accessible states, i.e., the states that the said system can be prepared into. However, it turns out that it is more suitable to look at anticommutators of states rather than their commutators. The reason is that the most basic quantum experiment is to interfere two different states $\rho_1$ and $\rho_2$. It is well known~\cite{ekertetal, carteret, Sjoqvist} that the interference fringes depend on the quantity $\tr[\rho_1\rho_2]$, which is the same as half of the trace of the anticommutator of the two  states, $\frac{1}{2}\tr[\{\rho_1,\rho_2\}]$. The main result of our paper will be that it is always possible (except for very special cases) by repeated measurements of anticommutators (and only anticommutators) to reveal any underlying quantumness. A properly defined interferometer will allow to detect the quantumness encoded in a given system. Although we speak of the quantumness of a single physical system, in practice we will need two or more copies of the physical system to witness the quantumness, as is almost always the case for other approaches. Finally, we show that our  witness is deeply linked to quantum discord~\cite{arXiv:1112.6238}. Before we show this, we first set the scene by introducing a formal definition of quantumness.

\section{Quantumness}
\label{sec:qness}

We shall focus for simplicity on finite-dimensional systems and their algebra of observables $\mathcal{A}$~\cite{Araki, BratteliRobinson}. The following two statements are equivalent~\cite{AVR,APVR}: given any pair  $X, Y\in\mathcal{A}$,
\begin{align}
&  \mathcal{A} \; \textrm{is commutative:} \; [X,Y]=XY-YX=0 ; \label{classicalC} \\
&  X\geq 0, Y\geq 0 \; \longrightarrow \; \{X,Y\}= XY + YX \geq 0 . \label{classicalA} 
\end{align}
Therefore, if the symmetrized product \eqref{classicalA} of two positive observables can take negative values, the algebra of observables is non-Abelian and the system is quantum. This motivated the notion of ``quantumness witness"~\cite{AVR,APVR}, that was experimentally tested in~\cite{brida08,bridagisin}.

Quantum states are non-negative Hermitian operators. Therefore the  characterization~\eqref{classicalC}--\eqref{classicalA} can be extended to \textit{states} as well and with many far reaching consequences. This is the main objective of this work. Consider a protocol by which an observer \cO is given two states $\rho_1$ and $\rho_2$, without knowing whether they commute. Assume that \cO can only perform interference experiments and therefore can only extract information about \textit{anti}commutators. Can \cO understand that the two states do not commute (and therefore that the system is quantum)? Notice that the equivalence between \eqref{classicalC} and \eqref{classicalA} pertains to the \textit{whole} algebra and not to any given two operators. In other words, $[\rho_1,\rho_2]\neq 0$ does not imply in the least that $\{\rho_1,\rho_2\}$ can take negative value. Consider for example two qubits in states $\rho= \frac{1}{2} (\openone + \bm{x} \cdot \bm\sigma)$ and $\rho'= \frac{1}{2} (\openone + \bm{x}' \cdot \bm\sigma)$. Then  $\{\rho,\rho'\}$ is positive definite if $|\bm{x}|^2 + |\bm{x}^{\prime}|^2 \le 1 + (\bm{x} \cdot \bm{x}^{\prime})^2 $. When a qubit can be prepared in states $\rho$ and $\rho'$ that do not commute but satisfy this condition, the anticommutator (first-order interference experiment) does not bring quantumness to light.

However, \cO can perform interference experiments of \textit{any} order, so that \cO can in principle obtain information about  repeated measurements of any anticommutators, such as  $\{\rho_1,\rho_1\}, \{\rho_1,\rho_2\}, \{\{\rho_1,\rho_1\},\rho_2\}$ and any (arbitrary) order of nested anticommutators (see Fig.~\ref{swap} for a possible scheme). The main result of this Article is a Theorem that proves the following statement: given any two states of a system, it is possible to bring to light quantumness by \textit{only} looking at the available anticommutators. 

The scheme to be discussed in this Article is of general validity. However, it is best suited for two of states whose commutator is ``significantly" nonvanishing. Since states are normalized to one, this yields a natural scale for their commutator and provides quantitative meaning to the above expression ``significant." The commutator of two states can vanish when the states are either parallel (namely, they admit a common eigenbasis) or orthogonal (their spans do not overlap). In both cases their anticommutator can be of no help. In the former case the two states are ``classical" with respect to each other and no quantumness can be brought to light (their anticommutator being always positive). In the latter case interference will vanish (and so will the anticommutator). This motivates the question: what happens when the commutator is very small (in the aforementioned natural scale)?
One expects that in such a case to unearth quantumness (nonpositivity of the anticommutator) requires significant resources. We shall consider such subtle cases in the Appendices. The fact that states should neither be parallel nor orthogonal suggests that the ideas explored here could be related to unambiguous state discrimination \cite{unambiguous} and probabilistic cloning \cite{probclon}.

\section{Bringing quantumness to light by nested anticommutators}
\label{sec:qnesslight}
Our strategy will be the following. We shall first observe (Theorem \ref{thm:1}) that if one of the two states is pure, one anticommutator suffices to bring quantumness to light. Therefore pure states are privileged and presumably quantumness will be easier to detect for states that are ``close" to pure states. We shall then observe (Theorem \ref{thm:2}) that any mixed state, as long as its maximum eigenvalue is not degenerate, can be made arbitrarily close to a pure state by iterating anticommutators for a finite number of times. The proof is an application of familiar statistical-mechanical concepts. Finally, Theorem \ref{thm:3} will show that the anticommutator of any two states that are sufficiently close to (non-orthogonal, non-parallel) pure states is not positive definite.  Our result will be valid for qudits of any dimension.

\subsection{Pure-state quantumness}
\begin{theorem}\label{thm:1}
If a state is pure, $\rho_1=\rho_1^2=\ket{\psi}\bra{\psi}$, then its anticommutator with any other state $\rho_2$ is a non-negative definite operator if and only if $[\rho_1,\rho_2] = 0$.
\end{theorem}

We start with a Lemma.

\begin{lemma}\label{lemm:1}
If a pure state $\rho_1=\rho_1^2=\ket{\psi}\bra{\psi}$ and a mixed state $\rho_2$ have vanishing anticommutator then their commutator is also vanishing.
\end{lemma}

\Proofa{Lemma \ref{lemm:1}}
Let $\ket{\psi}=\sum_i f^*_i \ket{\phi_i}$, where $\{\ket{\phi_i}\}$ is the eigenbasis or $\rho_2=\sum_i \lambda_i \ket{\phi_i}\bra{\phi_i}$ with $\sum_i \ket{\phi_i}\bra{\phi_i}=\openone$, and $\{\lambda_i\}$ are the eigenvalues of $\rho_2$, satisfying $0\le\lambda_i\le 1$ and $\sum_i\lambda_i=1$. By direct computation we have 
\begin{gather}
\{\rho_1,\rho_2\}=
\sum_{ij}(\lambda_i f^*_j f_i + \lambda_j f^*_i f_j ) \ket{\phi_j}\bra{\phi_i} =0.
\end{gather}
Since each diagonal element is vanishing, this implies $2 \lambda_i |f_i|^2=0$. This means that either $\lambda_i$ is vanishing or $f_i=\braket{{\psi}|{\phi_i}}=0$, i.e., $\ket{\psi}$ is orthonormal to the support of $\rho_2$ and therefore $\rho_1\rho_2=0$. \qed

\bigskip
\Proofa{Theorem \ref{thm:1}}
The proof is trivial one way, since both $\rho_1$ and $\rho_2$ are positive. For the converse we start by observing that if $\{\rho_1,\rho_2\} = 0$ the Theorem holds due to the preceding Lemma. We therefore assume $\{\rho_1,\rho_2\} \neq 0$.
Its anticommutator with the pure state $\rho_1$ yields
\begin{gather}
\{\rho_1,\rho_2\}=\sum_i\lambda_i(f_i\ket{\psi}\bra{\phi_i}+f^*_i\ket{\phi_i}\bra{\psi}),
\end{gather}
where $f_i=\braket{{\psi}|{\phi_i}}$ and $\sum_i|f_i|^2=\sum_i\braket{{\psi}|{\phi_i}}\braket{{\phi_i}|{\psi}}=1$. We normalize this anticommutator as
\begin{gather}
\rho_{12}=\frac{\{\rho_1,\rho_2\}}{\tr[\{\rho_1,\rho_2\}]}
=\frac{\sum_i\lambda_i(f_i\ket{\psi}\bra{\phi_i}+f^*_i\ket{\phi_i}\bra{\psi})}{2\sum_i \lambda_i |f_i|^2}.
\end{gather}
Since $\rho_{12}$ is a Hermitian and unit-trace operator, its purity $\tr[\rho_{12}^2]$ must be less than 1 to satisfy the positivity condition.
If on the other hand the purity exceeds unity, $\rho_{12}$ is proved not to be non-negative definite. We get
\begin{gather}
\tr[\rho^2_{12}]
=\frac{(\sum_{i}\lambda_i|f_i|^2)^2+\sum_{i}\lambda_i^2|f_i|^2}{2(\sum_i \lambda_i |f_i|^2)^2},
\end{gather}
and hence, the positivity of $\{\rho_1,\rho_2\}$ is violated if $\tr[\rho_{12}^2] > 1$, namely, if
\begin{gather}
\sum_i\lambda_i^2|f_i|^2>\biggl(\sum_i\lambda_i|f_i|^2\biggr)^2.
\end{gather}
This is always true by virtue of the Cauchy-Schwarz inequality (recall that $\sum_i|f_i|^2=1$), except when $\lambda_if_i=\lambda f_i$ with a real number $\lambda$ for all $i$, i.e., except when $\tr[\rho_{12}^2] = 1$. On the other hand, $\tr[\rho_{12}^2]=1$ implies 
\begin{align}
[\rho_1,\rho_2]
&=\sum_i\lambda_i(f_i|\psi\rangle\langle\phi_i|-f_i^*|\phi_i\rangle\langle\psi|)
\nonumber\\
&=\lambda \sum_i(f_i|\psi\rangle\langle\phi_i|-f_i^*|\phi_i\rangle\langle\psi|)
=0.
\end{align}
\qed

\bigskip
The above Theorem is useful to understand where and how one should look for anticommutators that can take negative values. This will be the subject of Theorem \ref{thm:3} in the following.

\subsection{Amplification of purity}
We showed in the preceding subsection that purity is a resource to witness quantumness. Now we restate a well known result of statistical mechanics that allows for amplification of purity as a Theorem.

\begin{theorem}\label{thm:2}
If a density operator $\rho$ does not have a degeneracy in its maximum eigenvalue, then the normalized operator corresponding to $\rho^n$ approaches, as $n\rightarrow \infty$, the pure state of the eigenvector corresponding to the largest eigenvalue.
\end{theorem}

\Proof
Let us write state $\rho$ in its eigenbasis,
\begin{gather}
	\rho=\sum_i\lambda_i \ket{i}\bra{i},
\end{gather}
$\lambda_0$ being the largest eigenvalue, corresponding to (nondegenerate) state $|0\rangle$, with no loss of generality.
Since $\lambda_i\le1$ in general, $\lambda_i^n$ decays as $n\to\infty$, but $\lambda_0^n$ decays most slowly, and we end up with
\begin{gather}
\lim_{n\rightarrow\infty}	\frac{\rho^n}{\tr[\rho^n]}
	=\lim_{n\rightarrow\infty} \sum_i\frac{\lambda^n_i}{\sum_j\lambda^n_j} \ket{i}\bra{i}
	=\ket{0}\bra{0}.
\end{gather}
\qed

\bigskip
Instead of taking $n$ to infinity and bringing a given mixed state to a pure state, like in Theorem \ref{thm:2}, we may just take a finite number of iterations and bring the state close to the pure state. Given any two mixed states, we can take them both $\ep$-close to pure states, respectively.  The anticommutator of these two states will admit (at least) a negative eigenvalue. Below we bound $\ep$ based on how close these two states are to each other. For the moment, we assume that the maximum eigenvalues of both given states are not degenerate. Degenerate cases will be commented later.

Observe first that we can define the closest pure state $\ket{\psi}$ to an arbitrary state $\rho$, in the sense that $\bra{\psi}\rho\ket{\psi}$ is maximum among all pure states. In that case $\ket{\psi}$ is an eigenvector of $\rho$ with the largest eigenvalue and $\rho$ can be expressed as a convex sum
\begin{gather}
\label{eqn:convex}
\rho=\lambda\ket{\psi}\bra{\psi}+(1-\lambda)\eta ,
\end{gather}
with a density operator $\eta$ ($\eta\ge0$, $\tr[\eta]=1$) that is orthogonal to $\ket{\psi}$, i.e., $\eta\ket{\psi}=0$ ($\lambda$ being the largest eigenvalue of $\rho$).
We are ready to prove our central result.

\subsection{Mixed-state quantumness}
\begin{theorem}\label{thm:3}
Given two non-commuting mixed states $\rho_1$ and $\rho_2$  close to pure states $\ket{\psi_1}$ and $\ket{\psi_2}$, respectively:
\begin{gather}
\rho_i=(1-\epsilon_i)\ket{\psi_i}\bra{\psi_i}+\epsilon_i\eta_i\quad(i=1,2),
\label{rho12bis}
\end{gather}
with $| \braket{{\psi_1}|{\psi_2}}| \ne 0 , 1$ and $\eta_1 \ket{\psi_1} = \eta_2\ket{\psi_2}=0$. The anticommutator is not positive semi-definite $\{\rho_1,\rho_2\}\ngeq 0$, provided $\epsilon_1$ and $\epsilon_2$ are small enough to satisfy 
\begin{gather}
\epsilon_1g_1+\epsilon_2g_2<(1-|f|^2)/2,
\label{eqn:EpsCond}
\end{gather}
where $g_1=\bra{\psi_2}\eta_1\ket{\psi_2}$, $g_2=\bra{\psi_1}\eta_2\ket{\psi_1}$ and $f=\langle\psi_1|\psi_2\rangle$ $(|f| \simeq \sqrt{\tr[\rho_1\rho_2]}$ if $\epsilon_{1,2}$ are small enough).
\end{theorem}

\Proof
The anticommutator of $\rho_1$ and $\rho_2$ reads 
\begin{align}
\{\rho_1,\rho_2\}
={}&(1-\epsilon_1)(1-\epsilon_2)
(f\ket{\psi_1}\bra{\psi_2}+f^*\ket{\psi_2}\bra{\psi_1}) \nonumber\\
&{}+\epsilon_1(1-\epsilon_2)(\eta_1\ket{\psi_2}\bra{\psi_2}+\ket{\psi_2}\bra{\psi_2}\eta_1) \nonumber\\
&{}+\epsilon_2(1-\epsilon_1)(\eta_2\ket{\psi_1}\bra{\psi_1}+\ket{\psi_1}\bra{\psi_1}\eta_2) \nonumber\\
&{}+\epsilon_1\epsilon_2\{\eta_1,\eta_2\},
\label{eqn::rho12Thm3}
\end{align}
with its trace given by
\begin{align}
\tr[\{\rho_1,\rho_2\}]
={}&2(1-\epsilon_1)(1-\epsilon_2)|f|^2
+2\epsilon_1(1-\epsilon_2)g_1\nonumber\\
&
{}+2\epsilon_2(1-\epsilon_1)g_2
+\epsilon_1\epsilon_2\tr[\{\eta_1,\eta_2\}].
\end{align}
Let $\rho_{12}={\{\rho_1,\rho_2\}}/{\tr[\{\rho_1,\rho_2\}]}$. This is a Hermitian and unit-trace operator, whose purity $\tr[\rho_{12}^2]$ must be less than 1 to satisfy the positivity condition. The anticommutator is therefore proved to be not positive definite, $\{\rho_1,\rho_2\}\ngeq0$, if the purity of $\rho_{12}$ exceeds unity.
This quantity is readily calculated, at first order in $\epsilon_1$ and $\epsilon_2$:
\begin{gather}
\tr[\rho_{12}^2]
=\frac{(1-2\epsilon_1-2\epsilon_2)(1+|f|^2)+2\epsilon_1g_1+2\epsilon_2g_2}{2[(1-2\epsilon_1-2\epsilon_2)|f|^2+2\epsilon_1g_1+2\epsilon_2g_2]},
\label{eqn:trfirstorder}
\end{gather}
and we get $\tr[\rho_{12}]^2>1$ under condition \eqref{eqn:EpsCond}. \qed

\begin{figure}[t]
\includegraphics[width=8cm]{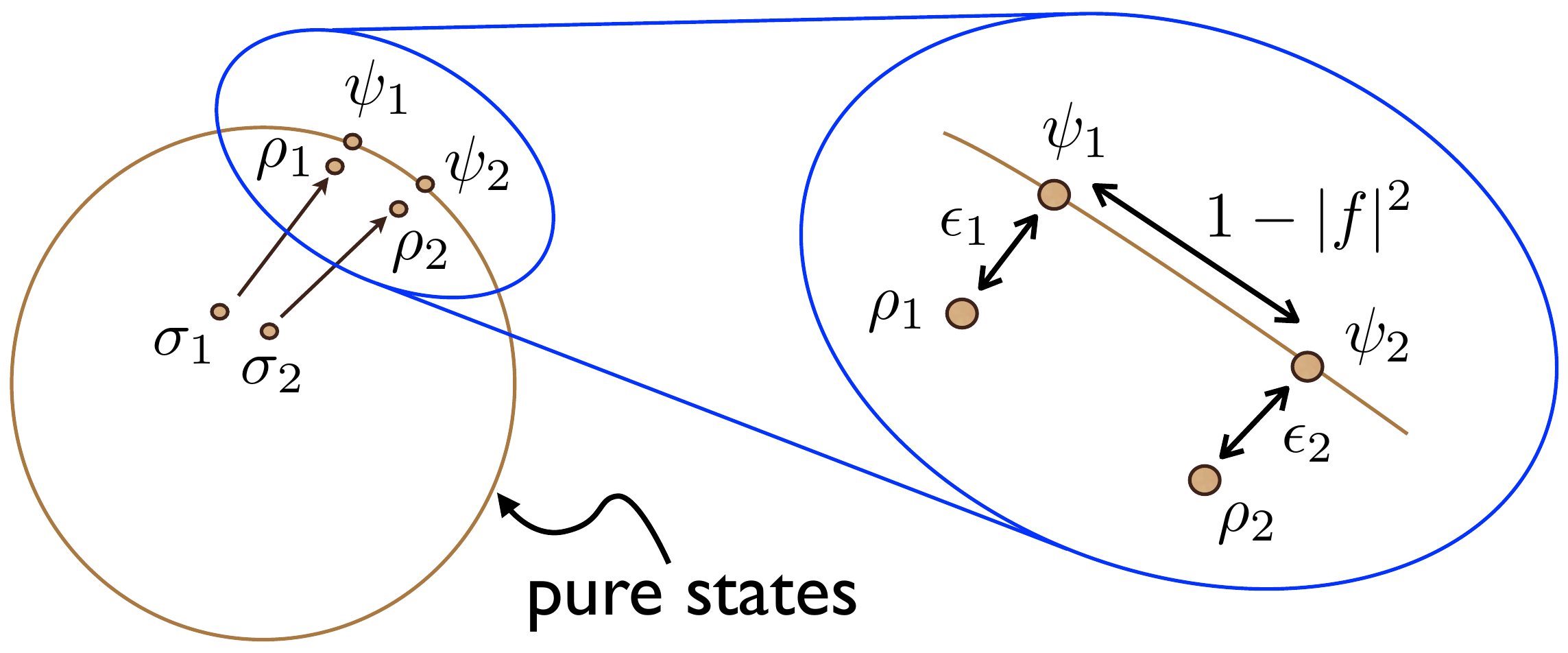}
\caption{Outline of the strategy to bring quantumness to light.
The circle pictorially represents the space of quantum states, and the states on the surface are pure states.
One starts with two (possibly highly) mixed states $\sigma_1$ and $\sigma_2$, that do not commute, and ``purifies" them (according to Theorem \ref{thm:2}) into states  $\rho_1 \propto \sigma_1^m$ and $\rho_2 \propto \sigma_2^n$, that are $\epsilon$-close to pure states $\psi_1$ and $\psi_2$, respectively.
Notice that this can be done in a finite number of steps ($m$ and $n$, respectively).
Then, their anticommutator $\{\rho_1,\rho_2\}$ jumps out of the state space, admitting a negative eigenvalue, under condition \eqref{eqn:EpsCond} of Theorem \ref{thm:3}, and witnesses quantumness.
Products of powers of density matrices (as well as their anticommutators) are associated to quantum interferometric circuits (see e.g., Fig.~\ref{swap}), so the whole strategy is based on quantum interference.
According to condition \eqref{eqn:EpsCond}, the larger the overlap $|f|^2=|\langle\psi_1|\psi_2\rangle|^2 \simeq \tr[\rho_1\rho_2]$ between the two states, the more difficult it is to show that anticommutator $\{\rho_1,\rho_2\}$ can take negative values, and therefore to witness quantumness.}\label{theorems}
\end{figure}

We are now in a position to put the results above in perspective. We have given a method to test for quantumness of a system that is preparable in two noncommuting states. However, if these states are sufficiently mixed then we simply take many copies of these states. According to Theorem \ref{thm:2} many copies simulate higher purity. Once the purity is high enough, according to Theorem \ref{thm:3}, we can find the nonpositive anticommutator. The strategy adopted in Theorems \ref{thm:2} and \ref{thm:3} in order to prove that the quantumness witness $\{\rho_1,\rho_2\}$ can take negative values is pictorially represented in Fig.~\ref{theorems}.

Technically, given states $\sigma_1$ and $\sigma_2$, that are neither parallel nor orthogonal, we may have to purify them in some finite rounds taking the anticommutator of each state with itself, which will purify the state due to Theorem \ref{thm:2}.
Once we attain desired purification, we can take the anticommutator of the two purified states.
Mathematically we can say that given two noncommuting states $\sigma_1$ and $\sigma_2$ we have a nonpositive operator corresponding to a nested 	anticommutator: $\{\rho_1, \rho_2\} = \{\sigma_1^m, \sigma_2^n\} \propto \{ \{\sigma_1, \dots \{\sigma_1, \sigma_1\} \}, \{\sigma_2, \dots \{\sigma_2, \sigma_2 \} \} \}$. 
The positivity of the operator on the left and on the right are the same. 
The equality is lacking only due to different normalization on the two sides. 
The values of $m$ and $n$ can be interpreted as the number of copies of the states $\rho_1$ and $\rho_2$, respectively, that are needed to witness the quantum feature of the system. This scheme is depicted in Fig.~\ref{theorems}. Note that $m=n=1$ when one of the states is pure due to Theorem \ref{thm:1}. Conversely, note that for highly mixed states $m$ and $n$ take large values to witness quantumness. For a related study on this topic see~\cite{MFPVY,FPVY}.

A few exceptional cases should be noted for concreteness.
In Theorem \ref{thm:3} we assume $|f|=|\braket{{\psi_1}|{\psi_2}}| \neq 0 , 1$. Orthogonal states do not interfere, their anticommutator $\{\ket{\psi_1} \bra{\psi_1}, \ket{\psi_2} \bra{\psi_2}\}$ vanishes, and does not detect quantumness. When the (mixed) states $\rho_1$ and $\rho_2$ of Theorem \ref{thm:3} are close to orthogonal states $\psi_1$ and $\psi_2$, as in \eqref{rho12bis}, their anticommutator almost vanishes (it is of order $\epsilon$), but it is still possible to bring quantumness to light. See Appendix \ref{app:f0}. On the other hand, notice that condition \eqref{eqn:EpsCond} becomes more and more stringent when $|f|^2$ is closer to unity, namely, when the two states $\ket{\psi_1}$ and $\ket{\psi_2}$ are very close to each other. When $|f|^2=1$, condition \eqref{eqn:EpsCond} is not valid. Note that in such a case $g_1=g_2=0$, and one must look at second-order terms in \eqref{eqn:trfirstorder}.
We then realize that it is not possible to bring quntumness to light when $|f|=1$. See Appendix \ref{app:f1}.
Another delicate situation occurs when the maximum eigenvalues of both $\rho_1$ and $\rho_2$ are degenerate.
For such cases, see Appendix \ref{degCase}.

\section{Measuring anticommutator}
\label{sec:measanti}
\begin{figure}[t]
\includegraphics[width=8cm]{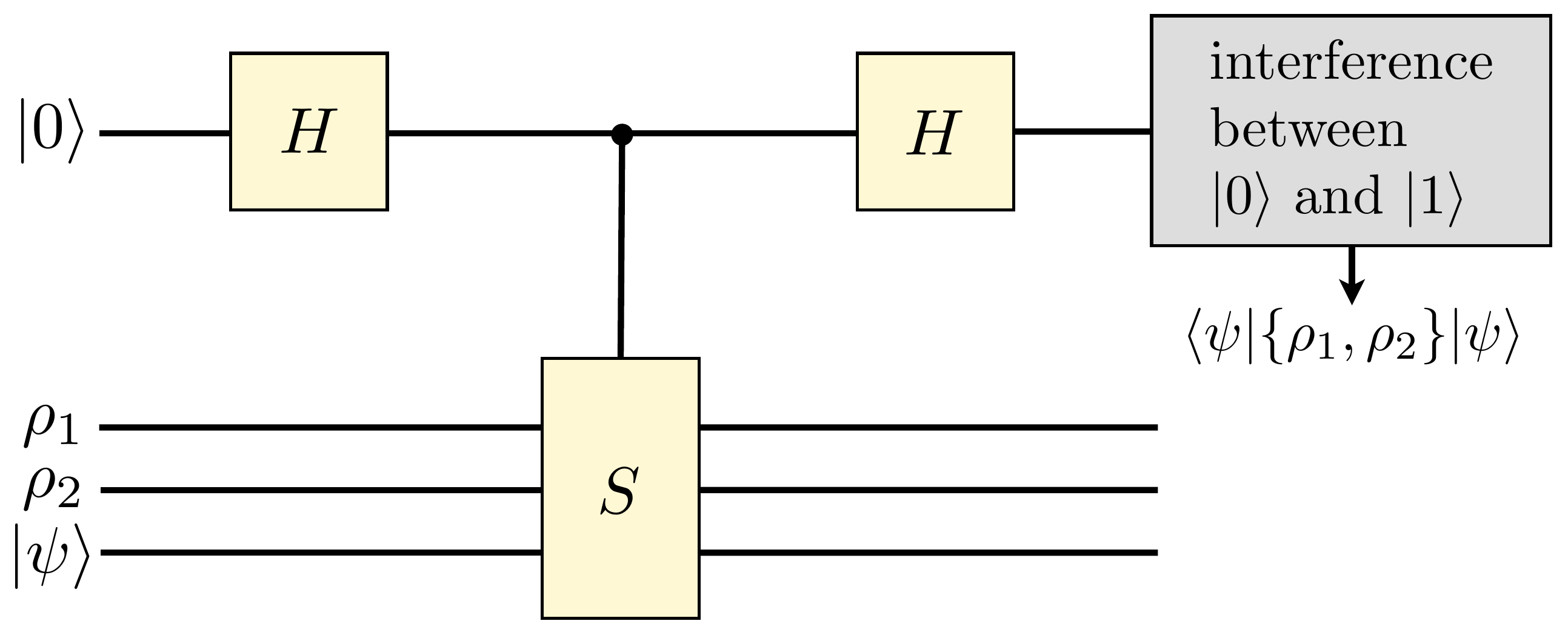} 
\caption{\label{swap} 
The strategy outlined in Theorems \ref{thm:2} and \ref{thm:3} and pictorially represented in Fig.\ \ref{theorems} hinges upon measurements of products of density matrices and their anticommutators.
These are nothing but interference experiments \cite{PhysRevA.64.052311,ekertetal, vlatkoqfi}.
We give here one such example.
$H$ is the Hadamard gate and $S$ is the \textsc{shift} gate.
If we choose $\ket{\psi}$ to be the eigenvector of $\{\rho_1,\rho_2\}$ corresponding to the nonpositive eigenvalue of $\{\rho_1,\rho_2\}$, then $\braket{\sigma_z}=\frac{1}{2}\braket{{\psi} |{ \{\rho_1,\rho_2\} }|{ \psi}}$ at the output port of the control qubit will show the quantumness we seek.}
\end{figure}

Our scheme hinges upon products of density matrices and their anticommutators. The traces of them can be measured by interference experiments, e.g., by the circuit given in Fig.~\ref{swap}, which involves a \textsc{shift} operator $S$~\cite{PhysRevA.64.052311,ekertetal, vlatkoqfi}, whose action is defined by 
\begin{gather}
\label{eqn:shift}
S\ket{ \psi_1, \psi_2, \dots, \psi_{l-1}, \psi_{l}} = \ket{ \psi_l, \psi_1, \psi_2, \dots, \psi_{l-1}}. 
\end{gather}
Notice that the trace of the \textsc{shift} operator's action from one side only yields
\begin{gather}
\label{eqn:swapn}
\tr[S(\rho_1 \otimes \rho_2 \otimes \dots \otimes \rho_l)] = \tr[\rho_1 \rho_2 \cdots\rho_{l}].
\end{gather}
Using a control qubit and implementing the controlled-\textsc{shift} operator, $\mathcal{C}_{S} = \ket{0} \bra{0} \otimes \openone + \ket{1} \bra{1} \otimes {S}$, we can measure the trace of the product of any number of density operators.

Let us sketch the main idea. 
The circuit shown in Fig.~\ref{swap} essentially represents a Mach-Zehnder interferometer for the control qubit (initially set in state $\ket{0}$). The Hadamard gate acts as a ``beam splitter" and yields the superposition $(\ket{0}+\ket{1})/\sqrt{2}$, where the two states $\ket{0}$ and $\ket{1}$ can be thought of as the two paths in the interferometer.
The phase difference between the two paths depends on the \textsc{shift} gate. 
The two beams are finally recombined at the second Hadamard gate, interfering with each other, and the difference between the probabilities of finding the control qubit in $\ket{0}$ and $\ket{1}$, which is the expectation value of $\sigma_z$ of the control qubit and is related to the visibility of the interference, reads
\begin{align}
\langle\sigma_z\rangle ={}& \frac{1}{2}\tr[S(\rho_1\otimes\rho_2\otimes\ket{\psi}\bra{\psi}) + (\rho_1 \otimes \rho_2 \otimes \ket{\psi}\bra{\psi}) S^\dag]   \nonumber \\
  ={}& \frac{1}{2}\bra{\psi} \{ \rho_1,\rho_2\}\ket{\psi},
\label{eqn:vis}
 \end{align}
due to formula (\ref{eqn:swapn}). Note that, if $\{\rho_1,\rho_2\}$ is not positive definite, there certainly exists a state $\ket{\psi}$ such that $\braket{{\psi} |{ \{\rho_1,\rho_2\}} |{\psi}}=q < 0$. This state can simply be taken to be the eigenvector of $\{\rho_1,\rho_2\}$ corresponding to (one of) its nonpositive eigenvalue(s). 

\section{Quantum discord}\label{sec:discord}
The notion of quantumness (of a single system) presented in this Article is related to quantum discord. Quantum discord and related measures~\cite{arXiv:1112.6238} attempt to quantify the quantum correlations in multipartite quantum states. However, for simplicity, we only work with bipartite states $\rho_{AB}$ here. Let us denote quantum discord (as measured by $B$) in $\rho_{AB}$ as $D(A|B)$.

Suppose Alice and Bob share a state and Alice has to convince (unaware) Bob that he is quantum correlated to her. She can do this by making two measurement of her system such that the corresponding conditional states of $B$ do not commute. Then she simply has to communicate the outcomes of her measurements and Bob can carry out the interference experiment, i.e., measure the nonpositivity of the anticommutator of the two states prepared by Alice. 
However, Alice can only prepare conditional states of Bob that do not commute if he is quantum correlated to her. By contrast, if Alice and Bob share a state with no quantum correlations, Alice can only prepare conditional states of Bob that commute. This comes from the following Theorem.

\begin{theorem}{[Chen \textit{et~al.}~\cite{PhysRevA.83.020101}]}
A bipartite state is quantum correlated for $B$ if and only if for any set of location operation by $A$, the conditional state of $B$ all commute:
\begin{gather}
D(A|B)=0 \quad \Longleftrightarrow \quad [\rho_{B|i},\rho_{B|j}]=0, \quad \forall i,j,
\end{gather}
where $\rho_{B|i}=\tr_A[\Lambda^i_A \otimes \mathcal{I}_B (\rho_{AB})]$ are the conditional states of $B$ for a (generalized) quantum operations $\Lambda^i_A$ that is made on $A$. Conversely, if $D(A|B)>0$, then $A$ can make two local operations yielding two conditional states for $B$ that do not commute. 
\end{theorem}

Once Alice remotely prepares two nonorthogonal states for Bob, he can measure the anticommutator of these states. In other words, carry out the procedure of Theorems \ref{thm:1}--\ref{thm:3} above. The anticommutator will be nonpositive if and only if Bob was quantum correlated to Alice. However the number of states that Alice has to produce for Bob depends on the mixedness of the conditional states and how noncommuting they are.

\section{Conclusions}
\label{sec:concl}
We have introduced a method to witness the quantumness of a system that is preparable in noncommuting states. The method relies on the fact that the anticommutator of two classical states is always positive. We show that there is always a nonpositive anticommutator due to any two noncommuting states. However, the positivity of the anticommutator is dependent on the purity of the states. In general, for highly mixed states we require many copies of the two states (or alternatively high-order interference) in order to witness quantumness.  On the other hand, detecting the witness remains difficult because it requires interacting many copies of the system (the coherent interaction is the controlled-\textsc{shift} operator in Fig.~\ref{swap}). Therefore, in the end, the scheme presented here is in agreement with the overwhelming lack of quantumness in the macroscopic world. It says that a macroscopic object is indeed quantum and one can even witness this quantumness, provided enough copies are available and a suitable apparatus that can (coherently) interact these objects. According to this scheme the lack of quantumness in the macroscopic world is due to the limitations on coherent interactions of large number of macroscopic systems. Lastly, we have linked this notion of quantumness to quantum correlation as quantified by quantum discord.

\acknowledgements
We thank Yu-Chun Wu for helpful comments. KM and VV acknowledge the support of the John Templeton Foundation. KM, RF, and VV acknowledge the financial support by the National Research Foundation and the Ministry of Education of Singapore. SP and KY would like to thank the Centre for Quantum Technologies of National University of Singapore for the hospitality. SP and KY are partially supported by the Joint Italian-Japanese Laboratory on ``Quantum Technologies: Information, Communication and Computation'' of the Italian Ministry for Foreign Affairs. SP is supported by PRIN 2010LLKJBX on ``Collective quantum phenomena: from strongly correlated systems to quantum simulators". KY is supported by the Grant-in-Aid for Young Scientists (B) (No.\ 21740294) and the Grant for Excellent Graduate Schools both from the Ministry of Education, Culture, Sports, Science and Technology (MEXT), Japan, and by a Waseda University Grant for Special Research Projects (2012A-878).

\appendix
\section{Theorem \ref{thm:3} for $|f|=0,1$} \label{sec:append}
In Theorem \ref{thm:3}, it is proved that the anticommutator $\{ \rho_1, \rho_2\}$ is not positive semi-definite if (\ref{eqn:EpsCond}) holds for the states $\rho_1$ and $\rho_2$ that are $\epsilon$-close to pure states $\ket{\psi_1}$ and $\ket{\psi_2}$ as (\ref{rho12bis}), respectively, with $|f|=|{\langle\psi_1|\psi_2\rangle}|\neq 0,1$.
In this Appendix, we look at the cases $|f|=0$ and $1$.

\subsection{The $|f|=0$ case} 
\label{app:f0}
If $f=\braket{{\psi_1}|{\psi_2}}=0$, the anticommutator in (\ref{eqn::rho12Thm3}) is reduced to
\begin{align}
\{\rho_1,\rho_2\}
={}&\epsilon_1(1-\epsilon_2)(\eta_1\ket{\psi_2}\bra{\psi_2}+\ket{\psi_2}\bra{\psi_2}\eta_1) \nonumber\\
&{}+\epsilon_2(1-\epsilon_1)(\eta_2\ket{\psi_1}\bra{\psi_1}+\ket{\psi_1}\bra{\psi_1}\eta_2) \nonumber\\
&{}+\epsilon_1\epsilon_2\{\eta_1,\eta_2\}.
\end{align}
Then, one gets, up to the second order in $\epsilon$,
\begin{gather}
\tr[\{\rho_1,\rho_2\}]^2
=4(\epsilon_1g_1+\epsilon_2g_2)^2+O(\epsilon^3)
\end{gather}
and
\begin{align}
\tr[\{\rho_1,\rho_2\}^2]
={}&2\epsilon_1^2(g_1^2+\bra{\psi_2}\eta_1^2\ket{\psi_2})
\nonumber\\
&{}+2\epsilon_2^2(g_2^2+\bra{\psi_1}\eta_2^2\ket{\psi_1})
+O(\epsilon^3),
\end{align}
where $g_{1,2}$ are defined below (\ref{eqn:EpsCond}).
The anticommutator $\{\rho_1,\rho_2\}$ is not positive semi-definite if
\begin{align}
&\tr[\{\rho_1,\rho_2\}^2]
-\tr[\{\rho_1,\rho_2\}]^2
\nonumber\\
&\quad\ \ %
=2\epsilon_1^2(\Delta\eta_1)_2^2
+2\epsilon_2^2(\Delta\eta_2)_1^2
-8\epsilon_1\epsilon_2g_1g_2
+O(\epsilon^3)
>0
\end{align}
with $(\Delta\eta_1)_2^2=\bra{\psi_2}\eta_1^2\ket{\psi_2}-g_1^2$ and $(\Delta\eta_2)_1^2=\bra{\psi_1}\eta_2^2\ket{\psi_1}-g_2^2$.
This condition holds true if $(\Delta\eta_1)_2^2(\Delta\eta_2)_1^2>4g_1^2g_2^2$;  otherwise, it can be fulfilled, e.g., by taking $\epsilon_2$ small enough to satisfy
\begin{gather}
\epsilon_2/\epsilon_1
<\frac{2g_1g_2
-\sqrt{4g_1^2g_2^2-(\Delta\eta_1)_2^2(\Delta\eta_2)_1^2}
}{(\Delta\eta_2)_1^2},
\end{gather}
by iterating the purification procedure for $\rho_2$.

\subsection{The $|f|=1$ case}
\label{app:f1}
Note first that when $f=\braket{{\psi_1}|{\psi_2}}=e^{i\chi}$ we also have $\eta_1\ket{\psi_2}=\eta_2\ket{\psi_1}=0$.
The anticommutator in (\ref{eqn::rho12Thm3}) is reduced to
\begin{align}
\{\rho_1,\rho_2\}
={}&(1-\epsilon_1)(1-\epsilon_2)
(e^{i\chi}\ket{\psi_1}\bra{\psi_2}+e^{-i\chi}\ket{\psi_2}\bra{\psi_1}) \nonumber\\
&{}+\epsilon_1\epsilon_2\{\eta_1,\eta_2\}.
\end{align}
In this case one gets
\begin{align}
\tr[\{\rho_1,\rho_2\}]^2
={}&4(1-\epsilon_1)^2(1-\epsilon_2)^2
+\epsilon_1^2\epsilon_2^2\tr[\{\eta_1,\eta_2\}]^2
\nonumber\\
&{}+4\epsilon_1\epsilon_2(1-\epsilon_1)(1-\epsilon_2)\tr[\{\eta_1,\eta_2\}]
\end{align}
and
\begin{gather}
\tr[\{\rho_1,\rho_2\}^2]
=4(1-\epsilon_1)^2(1-\epsilon_2)^2
+\epsilon_1^2\epsilon_2^2\tr[\{\eta_1,\eta_2\}^2].
\end{gather}
Taking the difference we have
\begin{align}
&\tr[\{\rho_1,\rho_2\}^2]
-\tr[\{\rho_1,\rho_2\}]^2
\nonumber\\
&\quad\ \ %
{}=-4\epsilon_1\epsilon_2\tr[\{\eta_1,\eta_2\}]+O(\epsilon^3),
\end{align}
which is always nonpositive for small $\epsilon$, since $\tr[\{\eta_1,\eta_2\}]\ge0$, and the quantumness cannot be brought to light.

\section{Degenerate case}\label{degCase}
The scheme outlined in Theorems 2 and 3 is not efficient if the maximum eigenvalues of both $\rho_1$ and $\rho_2$ are degenerate. 
Suppose that the largest eigenvalue of $\rho_i$ ($i=1,2$) is $d_i$-fold degenerate, i.e., 
\begin{gather}
\rho_i=(1-\epsilon_i)\frac{1}{d_i}P_i
+\epsilon_i\eta_i,
\end{gather}
with $P_i$ being a $d_i$-dimensional projector and $\eta_i$ a density operator such that $P_i\eta_i=\eta_iP_i=0$.
Then,
\begin{widetext}
\begin{equation}
\tr[\{\rho_1,\rho_2\}^2]
-(\tr\{\rho_1,\rho_2\})^2
=2(1-2\epsilon_1-2\epsilon_2)\frac{1}{d_1^2d_2^2}
\,\Bigl(
\tr[(P_1P_2)^2]
+\tr[P_1P_2]
-2(\tr[P_1P_2])^2
\Bigr)
+O(\epsilon).
\label{eqn:DegIneq}
\end{equation}
\end{widetext}
If $d_1=1$ or $d_2=1$, i.e., if the maximum eigenvalue of one of the two states $\rho_1$ and $\rho_2$ is not degenerate, and $[P_1,P_2]\neq0$, (\ref{eqn:DegIneq}) is positive definite for small $\epsilon$, and the quantumness can be brought to light.
This is a generalization of Theorem \ref{thm:1}.
On the other hand, if $d_1>1$ and $d_2>1$, i.e., the maximum eigenvalues of both $\rho_1$ and $\rho_2$ are degenerate, the sign of (\ref{eqn:DegIneq}) becomes undetermined.

Let us look at an example: take two projectors of a three-level system
\begin{gather}
P_1 = \begin{pmatrix} 1&0&0\\ 0&1&0\\ 0&0&0 \end{pmatrix} \quad \mbox{and} \quad
P_2 = \begin{pmatrix} 0&0&0\\ 0&1&0\\ 0&0&1 \end{pmatrix}.
\end{gather}
We define two states $\rho_1=\frac{1}{2}P_1$ and $\rho_2 = \frac{1}{2} R P_2 R^\mathsf{T}$, where
\begin{gather}
R = \begin{pmatrix} \cos\theta&0&-\sin\theta\\ 0&1&0\\ \sin\theta&0&\cos\theta \end{pmatrix}.
\end{gather}
In this case, an eigenvalue of the anticommutator $\{\rho_1,\rho_2\}$ is negative semi-definite.
However, one gets
\begin{align}
&\tr[(P_1P_2)^2]
+\tr[P_1P_2]
-2(\tr[P_1P_2])^2
\nonumber\\
&\qquad\qquad\qquad\qquad\qquad
=-(3+\sin^2\!\theta)\sin^2\!\theta,
\end{align}
which is negative semi-definite, and the quantumness cannot be brought to light.
Notice that, even if the purity of the normalized anticommutator of $\rho_1$ and $\rho_2$ is bounded by $1$, it does not imply that the anticommutator $\{\rho_1,\rho_2\}$ is positive semi-definite and can have a negative eigenvalue.

\end{document}